\documentclass{raa}%

\usepackage{graphicx}
\usepackage{multirow}

\title{Focal Surface Attitude Detection for LAMOST}

\author[T.-Z.Hu et al.]{T.-Z Hu$^{1,2,3}$, Y.Zhang$^{1,2}$\thanks{email:yzh@niaot.ac.cn},X.-Q.Cui$^{1,2}$\thanks{email:xcui@niaot.ac.cn},  Y.-P.Li$^{1,2}$, X.-S.Pan$^{1,2,3}$, Y.Fu$^{1,2,3}$
\affil{$^1$National Astronomical Observatories/Nanjing Institute of Astronomical Optics \& Technology, Chinese Academy of Science, Nanjing 210042, China}%
\affil{$^2$CAS Key Laboratory of Astronomical Optics \& Technology, Nanjing Institute of Astronomical Optics \& Technology, Nanjing 210042, China}%
\affil{$^3$University of the Chinese Academy of Sciences, Beijing 10049, China}%
}%

\doi{10.1017/pas.\the\year.xxx}
\jyear{\the\year}

\usepackage{aas_macros}
\usepackage{hyperref} 
\hypersetup{colorlinks, citecolor=blue, linkcolor=blue, urlcolor=blue}

\begin{document}

\begin{frontmatter}
\maketitle

\begin{abstract}
With telescope apertures becoming larger and larger, the deployment of large-field telescopes is becoming increasingly popular. However, optical path calibration is necessary to ensure the image quality of large-field and large-diameter telescopes. In particular, focal plane attitude calibration is an essential optical path calibration technique that has a direct impact on image quality. In this paper, a focal plane attitude detection method using eight acquisition cameras is proposed based on the calibration requirements of the wide-field telescope, LAMOST. Comparison of simulation and experimental results shows that the detection accuracy of the proposed method can reach 30 arcsec. With additional testing and verification, this method could be used to facilitate regular focal plane attitude calibration for LAMOST as well as other large-field telescopes.

\end{abstract}

\end{frontmatter}

\section{INTRODUCTION }
\label{sec:intro}

A telescope is an essential tool for exploring the universe. To achieve highly accurate celestial body information, image quality is especially crucial. The observatory site, aperture, and optical path calibration have a significant impact on the image quality of the telescope. Focal plane attitude calibration, as an element of optical path calibration, is particularly important\citep{mcleod1996collimation,schechter2011generic}. The condition of the focal plane attitude directly affects the shape of the image.

The Large Sky Area Multi-Object Fiber Spectroscopic Telescope (LAMOST) is the largest-diameter and highest-spectral-acquisition-rate optical telescope with wide field of view (FOV). The telescope consists of a reflective, active, aspheric Schmidt corrector plate (Ma); a spherical primary mirror (Mb); and a focal surface. The LAMOST telescope has a $5^\circ$ FOV, and the focal plane diameter is 1.75 m.\citep{cui2012large,zhao2012lamost} Deviation of the attitude from the ideal focal plane will cause image spot distortion, resulting in energy loss into the optic fibre. 

The image quality required is that 80\% of the optical energy should be within 1.5 arcsec \citep{su2004active}. According to simulation data, it will cause the image size of the LAMOST telescope to increase by 0.1 arc seconds if focal plane rotates around the horizon axis about 54 arc-seconds in the reverse direction or 2.4 arcmins in the forward direction or 2.3 arcmins around the vertical axis. To avoid unnecessary energy loss, focal plane attitude calibration is required regularly.

Large FOV telescopes currently built include the Sloan Digital Sky Survey(SDSS)\citep{york2000sloan}, VLT Survey Telescope (VST)\citep{shanks2015vlt}, Two Micron All-Sky Survey(2MASS) \citep{skrutskie2006two}, Visible and Infrared Survey Telescope for Astronomy(VISTA) \citep{sutherland2015visible}, and the Wide-field Infrared Survey Explorer(WISE) \citep{wright2010wide}. Upcoming construction includes the Large Synoptic Survey Telescope(LSST)\citep{tyson2002large} and the Wide Field Infrared Survey Telescope (WFIRST)\citep{content2013wide}. Focal plane attitude detection also is essential for the above telescopes.

Traditional focal plane attitude detection uses the self-collimation method\citep{hemayed2003survey,li2016high}, which requires active optical self-collimation technology. This approach is affected by atmospheric visibility and suffers from low detection accuracy. Some telescopes adopt alternative approaches, such as the focal surface camera used by the VISTA telescope of the European Southern Observatory. VISTA's focal surface camera captures bright star images, using image size to detect the focal plane attitude. This technique requires $180^\circ$ focal plane rotation and a large-area charged couple device (CCD).\citep{sutherland2015visible,terrett2010interaction,dalton2006vista} However, LAMOST is a spectral survey telescope with 4000 optical fibres. This method cannot be applied in LAMOST.

Aiming at the characteristics of the wide FOV of LAMOST, it is proposed to use the existing eight acquisition cameras on the focal plane to obtain a series of defocused spot images on either side of the ideal focus surface, and to calculate the focal plane attitude based on the defocus values from all of the cameras.

Section 1 of this paper introduces the principles of our detection method. The simulation experiment and error analysis presented in Section 2 verify the feasibility of our approach, while Section 3 introduces the experimental results of obtained using our method of detecting the focal plane attitude of the LAMOST telescope. In the final section, we summary summarise our work and look forward to discuss the application of our method in the future.
 
\section{Detection method}
\begin{figure*}[htb]
\begin{center}
\includegraphics[width=25pc, height=15pc]{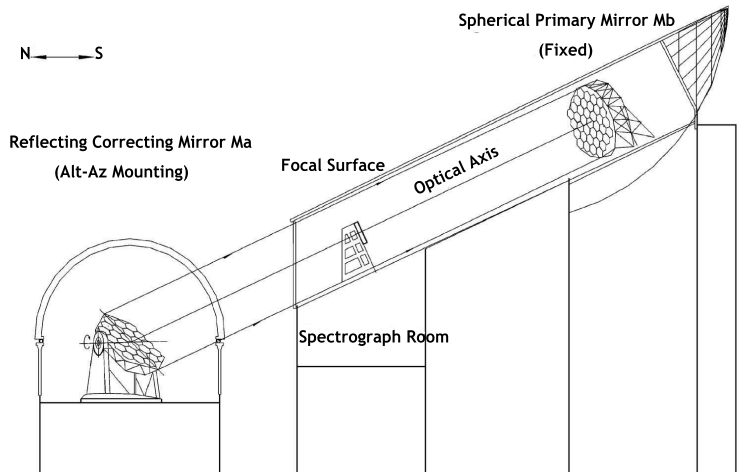}
\caption{LAMOST: a general view. LAMOST is a special reflecting Schmidt telescope with a 4 m aperture and a $5^\circ$ FOV. It has a focal length of 20 m and an f-ratio of 5. Its optical axis is fixed in the meridian plane and is tilted by $25^\circ$ to the horizontal. Celestial objects are observed for 1.5 hours as they cross the meridian. During the observation period only the alt-azimuth mount of the reflecting Schmidt correcting mirror (Ma) and the focal surface do the tracking. The survey area to be observed is $-10^\circ\leq\delta\leq+90^\circ$.}
 \label{Fig1}
\end{center}
\end{figure*}

\begin{figure}[htb]
\begin{center}
\includegraphics[width=\columnwidth]{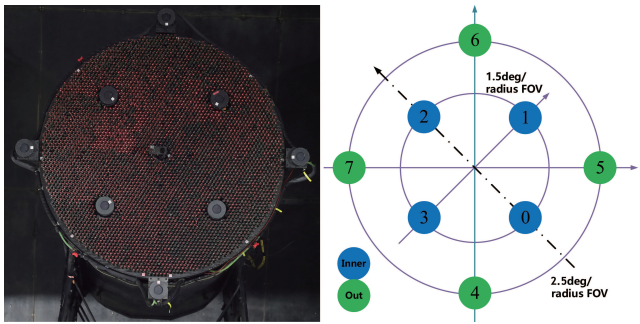}
\caption{The picture shows the focal surface of LAMOST; The left picture is the photo of the focal surface which is mounted with eight acquisition cameras, the right image shows the relative positions of eight cameras, four inner cameras in the 3-degree FOV in blue color and four outer cameras in the 5-degree FOV in green color.}\label{Fig2}
\end{center}
\end{figure}

The LAMOST optical system includes mirror Ma, mirror Mb, and focal surface, as shown in Fig.\ref{Fig1}. Ma ($5.72m \times 4.40m$) consists of 24 hexagonal segments with a diagonal of 1.1 m. Mb ($6.67 m \times 6.05 m$) has 37 hexagonal segments with a diagonal diameter of 1.1 m. The Mb has a radius of curvature of 40 m and a system focal length of 20 m \citep{su2012atmospheric}. Focal surface have 1.75 meters diameter and eight acquisition cameras, the focal surface picture and arrangement of eight acquisition camera are shown in Fig.\ref{Fig2}.

According to geometrical optics, the image has a pupil shape in the case of long-distance defocusing, and the image size increases linearly with the defocus distance \citep{born2013principles}.

Taking the positive direction of defocus (toward Ma) as an example, the position of the focus can be calculated by obtaining a series of defocus positions $x_n$ and the size $y_n$ of the defocused image 
$$
\begin{bmatrix}
	(x_1,y_1) & (x_2,y_2) & (x_3,y_3) & \cdots & (x_n,y_n) \\
\end{bmatrix}
$$

Because the image size increases linearly with the defocus distance, the relationship of image size and defocus distance satisfies Equation 1.

\begin{equation}
    \left[ \begin{array}{c}
        y_1 \\
    	y_2 \\
    	y_3 \\
    	\vdots \\
    	y_n \\
    \end{array}
    \right ]=a_1\times{
    \left [ \begin{array}{c}
	x_1-x_{01} \\
	x_2-x_{01} \\
	x_3-x_{01} \\
	\vdots \\
	x_n-x_{01} \\
	\end{array}
	\right ]}
\end{equation}
where $x_{01}$ is the theoretical focus position estimate value.

We can obtain the value of $a_1$ and $x_{01}$ by linear fitting with the least squares method. 

Furthermore, because of aberration, we need to consider the case of negative direction defocus (toward Mb) to get the estimated value $x_{02}$ of the theoretical focus position. The final focus position is given by the average of two independent $x$ values, $x_{03}=(x_{01}+x_{02})/2$, and the deviation of the focus position $\Delta f=x_{03}-x_{focal}$ is calculated where $x_{focal}$ is the ideal focus position.

In order to obtain the tilt angle $\Delta\theta$ of the focal surface relative to the two detector directions, we need to compare the relative deviation $\Delta{f^\prime}=\Delta{f_1}-\Delta{f_2}$ of the two detectors and relative distance $D$ between the two detectors.

\begin{equation}\label{eq1}
\Delta\theta=\Delta{f^\prime}/D,
\end{equation}

\begin{figure}[htb]
\begin{center}
\includegraphics[width=\columnwidth]{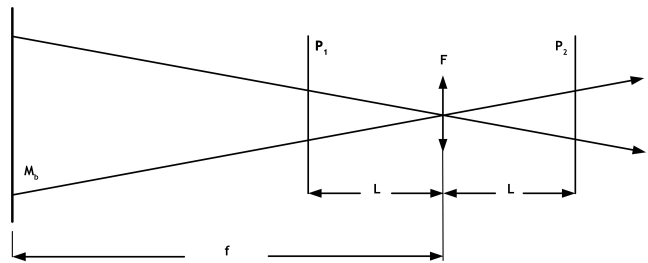}
\caption{Light reflected by the Mb mirror image onto the surface $F$. $P_1$ and $P_2$ are surfaces with the same defocus distance. The optic construction is similar to the curvature wavefront sensor, but our system moves the focal surface and variable defocus distance.}\label{Fig3}
\end{center}
\end{figure}

Further, before and after the focal surface, we explore the light intensity distribution of the image, as shown in Fig.\ref{Fig3}, where $P_1$ and $P_2$ are two surfaces with defocus distance $L$.

The distribution of light complex amplitude before passing reflector $M_b$ is 
\begin{equation}
	\psi(r)=A_0exp(i\frac{2\pi}{\lambda}w(r)) 
\end{equation}

According to the Fraunhofer diffraction approximation, the light complex amplitudes of the images on $P_1$ and $P_2$, respectively, are \citep{goodman2005introduction}

\begin{equation}
A_{P1}(r)=\psi(r)\otimes\{\frac{1}{i\lambda(f-l)}exp[i\pi\frac{r^2}{\lambda(f-l)}]\}\\
\end{equation}

\begin{equation}
A_{P2}(r)=\psi(r)\otimes\{\frac{1}{i\lambda(f+l)}exp[i\pi\frac{r^2}{\lambda(f+l)}]\}\\
\end{equation}

Obtained by geometric optical approximation, the relationship of the light intensities in the $P_1$ and $P_2$ planes is \citep{roddier1988curvature}
\begin{equation}
     \frac{CI_{P2}(r)-DI_{P1}(-r)}{CI_{P2}(r)+DI_{P1}(-r)}	=S
\end{equation}

\begin{equation}\label{eq2}
	S=\frac{C-D}{C+D}+\frac{2D}{C+D}\frac{f(f-l)}{l}[\bigtriangledown^2w(\frac{fr}{l})-\frac{\partial}{\partial n}w(\frac{fr}{l})\delta_c]
\end{equation}

where $I_{P1}(-r)$ and $I_{P2}(r)$ are the symmetric points light intensities of the $P_1$ and $P_2$ planes, respectively, and $C$, $D$ are constants, which satisfy
 
\begin{equation}
	\frac{D}{C}=(f+l)(f-l)
\end{equation}

It can be seen that the intensity distribution at the same distance before and after the focus is asymmetrical, unless wavefront curvature $\bigtriangledown^2w$ and wavefront radial tilt $\frac{\partial w}{\partial n}$ are zero. 

Not only does the light intensity distribution differ at a given distance before and after the focal surface, but the particular pupil shape creates difficulty in measuring the image sizes. A high-precision detection method is needed to extract the size of the defocus image.

In a comparison of the ellipse fitting algorithm \citep{fitzgibbon1999direct,gander1994least}, multi-peak Gaussian fitting algorithm \citep{guo2011simple,wan2018star}, average spacing algorithm (ASA), and peak spacing algorithm\citep{hu2020defocus}, the ASA has the highest detection accuracy and is least influenced by stellar brightness change. This algorithm can guarantee the accuracy of our image size detection. Details of the average spacing algorithm are given in Section 3.1. 

\section{Simulation and analysis}

Before the experiment, we used ZEMAX software for simulation analysis to verify the feasibility of our method \citep{geary2002introduction}. We built the LAMOST simulation software optical path to generate the defocusing images and calculated the defocus image sizes using the ASA.

\begin{figure}[htb]
\begin{center}
\includegraphics[width=\columnwidth]{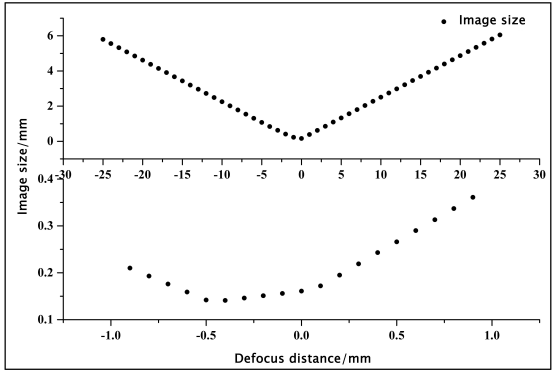}
\caption{The image size changes with the defocus distance. Near the focal surface, spot size change is small. On the 1 cm to 15 cm defocus distance scale, image size decreases with the defocus distance but not linearly. When the defocusing distance exceeds 15 cm, spot size decreases linearly with the defocus distance. The detail data of defocus and image size is shown in appendix A}\label{Fig4}
\end{center}
\end{figure}

Using the optical path of the LAMOST telescope, ray tracing can be used to obtain the theoretical spot size. When the star is near the culmination of the $30^\circ$ survey area, we detect the defocus image size from the negative defocusing 25 mm position to the positive defocusing 25 mm position and select the 100\% light energy-containing region as the spot size, as shown in Fig.\ref{Fig4}. 

We can see that the size of the image near the focus changes nonlinearly. To ensure linear behaviour, we chose defocus distances of 20-25 mm in the simulation analysis.

\subsection{Average spacing algorithm}

The ASA is used to estimate the image size, which is then simulated by the Zemax software. First, we choose a clear image and abstract the image area $k\times{k}$, which is an intensity distribution matrix $I$. $I(x,y)$ is light intensity in $(x,y)$. Second, we find the maximum light intensity matrix $A$ and minimum light intensity matrix $B$ of light intensity distribution matrix $I$. Values $A(x)$ and $B(x)$ are the maximum and minimum values of row $x$ of $I$, respectively. 

\begin{small}
\begin{equation}
    A={
    \left[ \begin{array}{c}
    	A(1)\\
    	A(2)\\
    	A(3)\\
    	\vdots \\
    	A(k)\\
    \end{array}
    \right ]}=max{
    \left [ \begin{array}{cccc}
	I(1,1) & I(1,2) &\cdots & I(1,k) \\
	I(2,1) & I(2,2) &\cdots & I(2,k) \\
	I(3,1) & I(3,2) &\cdots & I(3,k) \\
	\vdots & \vdots &\ddots & \vdots \\
	I(k,1) & I(k,2) &\cdots & I(k,k) \\
	\end{array}
	\right ]}
\end{equation}
\end{small}

\begin{small}
\begin{equation}
    B={
    \left[ \begin{array}{c}
        B(1)\\
    	B(2)\\
    	B(3)\\
    	\vdots \\
    	B(k)\\
    \end{array}
    \right ]}=min{
    \left [ \begin{array}{cccc}
	I(1,1) & I(1,2) &\cdots & I(1,k) \\
	I(2,1) & I(2,2) &\cdots & I(2,k) \\
	I(3,1) & I(3,2) &\cdots & I(3,k) \\
	\vdots & \vdots &\ddots & \vdots \\
	I(k,1) & I(k,2) &\cdots & I(k,k) \\
	\end{array}
	\right ]}
\end{equation}
\end{small}

Using average value $B_{ave}$ of $B$ as background noise, $C=A-B_{ave}$ represents new light intensity distribution. $I_{ave}$ is the average value of $C$. 

\begin{figure}[htb]
\begin{center}
\includegraphics[width=\columnwidth]{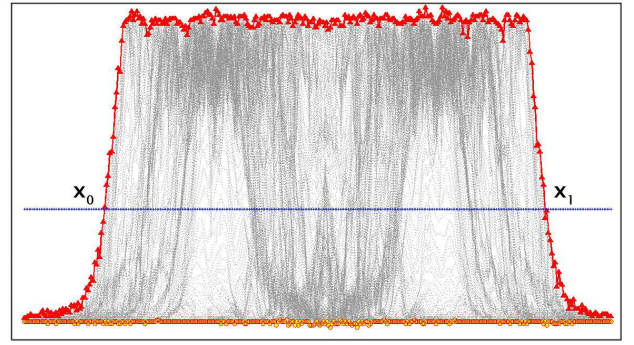}
\caption{The image process by the average spacing algorithm. Red $'\bigtriangleup '$ is maximum light intensity $A$, orange $'\circ '$ is minimum light intensity $B$, and the blue line is $a=0.5\times I_{ave}+B_{ave}$. $x_0$ and $x_1$ are the points of intersection of the blue and red lines.}
\label{Fig5}
\end{center}
\end{figure}

Using Lagrangian interpolation to improve detection accuracy, we find the values $x_0$ and $x_1$ that satisfy $A(x_0)=a$ and $A(x_1)=a$, where $a=0.5\times I_{ave}+B_{ave}$. $L=\lvert x_0-x_1 \lvert$ is the size of image Fig.\ref{Fig5}.

\subsection{Focal point and image size detection for constructing the look-up table}

\begin{figure}[htb]
\begin{center}
\includegraphics[width=\columnwidth]{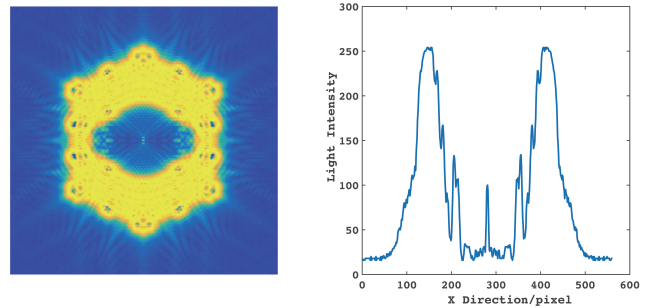}
\caption{At left, a simulation image by Zemax software. At right, the middle section light intensity distribution of the simulation image. The light intensity distribution is uneven because every segmented mirror produces a pupil image in the detector. Moreover, because of the vignette effect, some regions have low light intensity.}
\label{Fig6}
\end{center}
\end{figure}

Using a simulated optical path of a $30^\circ$ survey area with the central star in the 5-degree hour angle position before the culminant, we simulated the images at defocus distances of 20, 21, 22, 23, 24 and 25 mm in all eight acquisition cameras. We used the Zemax software and the ASA to calculate the image size, then evaluated the focus position. A simulation image of central field of view by Zemax software is shown in Fig.\ref{Fig6}. The test result of the eight fields of view, corresponding to LAMOST's actual eight acquisition cameras, is shown in Table.\ref{tab1}, Table.\ref{tab2}, Table.\ref{tab3}, and Table.\ref{tab4}.

\begin{table}[htb]
\caption{Image size estimated by Zemax software. Four FOVs corresponding to the four acquisition cameras in the inner circle of the focal plane (unit: mm).}
\centering
\begin{tabular}{@{}ccccc@{}}
\hline\hline
Distance & -1.5/-1.5 & 1.5/1.5 & 1.5/-1.5 & -1.5/1.5\\
\hline%
-25 & 5.914 & 5.804 & 5.794 & 6.027 \\
-24 & 5.671 & 5.577 & 5.558 & 5.793  \\
-23 & 5.429 & 5.350 & 5.322 & 5.558 \\
-22 & 5.187 & 5.123 & 5.086 & 5.324 \\
-21 & 4.944 & 4.896 & 4.850 & 5.089 \\
-20 & 4.702 & 4.669 & 4.615  & 4.855 \\
20 & 5.018 & 4.441 & 4.858 & 4.568 \\
21 & 5.261  & 4.668 & 5.094 & 4.802 \\
22 & 5.503 & 4.895 & 5.330 & 5.037 \\
23 & 5.745 & 5.122  & 5.566 & 5.271 \\
24 & 5.987 & 5.349  & 5.802 & 5.506 \\
25 & 6.230 & 5.576  &6.038  &5.741   \\
\hline\hline
\end{tabular}
\label{tab1}
\end{table}
\begin{table}[htb]
\caption{Image size estimated by Zemax software. Four FOVs corresponding to the four acquisition cameras in the outer circle of the focal plane (unit: mm).}
\centering
\begin{tabular}{@{}ccccc@{}}
\hline\hline
Distance & -2.5/0 & 2.5/0 & 0/-2.5 & 0/2.5\\
\hline%
-25 & 6.352 & 5.972 & 5.143 & 5.172 \\
-24 & 6.098 & 5.734 & 4.930 & 4.972 \\
-23 & 5.844 & 5.495 & 4.718 & 4.772 \\
-22 & 5.590 & 5.256 & 4.505 & 4.572 \\
-21 & 5.336 & 5.018 & 4.293 & 4.372 \\
-20 & 5.082 & 4.779 & 4.080 & 4.172 \\
20 & 5.137 & 4.793 & 4.430 & 3.838 \\
21 & 5.390 & 5.031 & 4.643 & 4.038 \\
22 & 5.644 & 5.270 & 4.855 & 4.238 \\
23 & 5.897 & 5.508 & 5.068 & 4.438 \\
24 & 6.150 & 5.747 & 5.281 & 4.638 \\
25 & 6.403 & 5.986 & 5.493 & 4.839 \\
\hline\hline
\end{tabular}
\label{tab2}
\end{table}

\begin{table}[htb]
\caption{Image size detected by the average spacing algorithm. Four FOVs corresponding to the four acquisition cameras in the inner circle of the focal plane (unit: pixel).}
\centering
\begin{tabular}{@{}ccccc@{}}
\hline\hline
Distance & -1.5/-1.5 & 1.5/1.5 & 1.5/-1.5 & -1.5/1.5\\
\hline%
-25 & 205.243 & 200.383 & 191.581 & 202.136 \\
-24 & 196.597 & 192.392 & 183.604 & 194.741 \\
-23 & 188.668 & 185.286 & 175.271 & 187.004 \\
-22 & 179.767 & 177.522 & 168.011 & 178.980 \\
-21 & 170.806 & 169.555 & 160.568 & 170.602 \\
-20 & 162.076 & 160.525 & 152.003 & 162.864 \\
20 & 171.494 & 149.638 & 158.530 & 152.273 \\
21 & 180.124 & 157.081 & 164.968 & 159.770 \\
22 & 188.921 & 165.174 & 172.971 & 167.813 \\
23 & 196.941 & 171.901 & 180.610 & 175.601 \\
24 & 205.350 & 180.154 & 188.978 & 183.309 \\
25 & 214.305 & 188.538 & 196.573 & 191.199 \\
\hline\hline
\end{tabular}
\label{tab3}
\end{table}

\begin{table}[htb]
\caption{Image size detected by the average spacing algorithm. Four FOVs corresponding to the four acquisition cameras in the outer circle of the focal plane (unit: pixel).}
\centering
\begin{tabular}{@{}ccccc@{}}
\hline\hline
Distance & -2.5/0 & 2.5/0 & 0/-2.5 & 0/2.5\\
\hline%
-25 & 289.992 & 243.324 & 241.658 & 239.024 \\
-24 & 278.401 & 233.247 & 231.249 & 230.774 \\
-23 & 267.659 & 222.918 & 221.831 & 220.711 \\
-22 & 256.843 & 213.327 & 212.379 & 211.821 \\
-21 & 245.448 & 202.002 & 201.395 & 201.511 \\
-20 & 233.306 & 192.126 & 192.480 & 193.500 \\
20 & 236.180 & 188.967 & 209.502 & 182.270 \\
21 & 247.978 & 198.930 & 220.209 & 191.721 \\
22 & 258.419 & 209.706 & 229.280 & 200.911 \\
23 & 269.939 & 220.155 & 239.908 & 210.524 \\
24 & 281.076 & 229.203 & 249.292 & 219.650 \\
25 & 292.098 & 239.463 & 259.509 & 229.589 \\
\hline\hline
\end{tabular}
\label{tab4}
\end{table}

The ideal focal position based on LAMOST's unique design is an average result derived from different survey areas \citep{cui2012large}. According to the results of the image size calculations using Zemax software and the ASA, the focal point estimation position using our method experienced have some deviation from the focal surface. Nevertheless, the estimate of focal position generated using Zemax software and the ASA is close enough to focal surface to construct a look-up table \citep{lucente1993interactive}for experimental testing. 

\begin{table}[htb]
\caption{Focal point evaluation and deviation value using Zemax software and the average spacing algorithm (ASA) for eight fields of view. A negative value indicates evaluation at the Mb side and a positive value indicates evaluation at the Ma side. The results reveal that the ASA and Zemax software produce similar results, and the ASA can be used to evaluate the ideal focal point position (unit: $\mu m$).} 
\centering
\begin{tabular}{@{}ccccc@{}}
\hline\hline
Field & Zemax software & ASA & diviation \\
\hline%
-1.5/-1.5 & -654.6 & -684.6 & -30.0  \\
1.5/1.5   & 501.0  & 571.7  & 70.7   \\
1.5/-1.5  & -516.0 & -519.1 & -3.1   \\
-1.5/1.5  & 619.4  & 549.6  & -70.2  \\
-2.5/0    & -139.3 & -185.6 & -46.3  \\
2.5/0     & -29.6  & 2.2    & 31.8   \\
0/-2.5    & -819.1 & -799.5 & 20.4   \\
0/2.5     & 842.3  & 755.4  & -89.6  \\
\hline\hline
\end{tabular}
\label{tab5}
\end{table}

Comparing the evaluation value of the focus point calculated by our algorithm and the ZEMAX simulation, which is shown in Table.\ref{tab5}, the detection accuracy of the four detectors in the inner circle is within $70\mu m$ while the outer ring is within $90\mu m$. The detection accuracy of the tilt angle is $30^{\prime\prime}$. 

 When central star on the culminant the defocuses image size of different filed is shown in Table.\ref{tab13}. In the experiment, when we get the star image size of different to defocus distance, we using the linear fitting which shows a good linearly and Lagrangian interpolation method to process data and get the focus position of different cameras. We use look up table method to compare the focal point position of the different camera with the simulation result to get the defocus value of the different camera. Then according to formula 2, we can get the focal surface tilt value. 

\begin{table*}[htb]
\caption{The defocus image size and focus position when the central star in the culminant position. Position (0) is focal estimate position with the central star in the culminant position, position (5) is focal estimate position with a central star at 5-degree hour angle position before culminant. It can been seen that the left position and right position have good symmetry and have a similar focal position to 5-degree hour angle.(unit:mm)}
\centering
\begin{tabular}{@{}ccccccccc@{}}
\hline\hline
DISTANCE/mm & 1.5/1.5 & 1.5/-1.5 &	-1.5/1.5 & -1.5/-1.5 & 2.5/0 & -2.5/0 &	0/2.5 &	0/-2.5\\
\hline%
20 & 4.5034 & 4.8744 & 4.5034 &	4.8744 &	 4.8391 & 4.8391 & 3.9066 &	4.3684\\
21 & 4.7338 & 5.1113 & 4.7338 & 5.1113 & 5.0789 & 5.0789 & 4.1105 & 4.5782\\
22 & 4.9642 & 5.3482 & 4.9642 & 5.3482 & 5.3187 & 5.3187	 & 4.3146 & 4.7882\\
23 & 5.1946 & 5.5851 & 5.1946 & 5.5851 & 5.5586 & 5.5586	 & 4.5187 & 4.9981\\
24 & 5.4252 & 5.8220 & 5.4252 & 5.8220 & 5.7985 & 5.7985	 & 4.7227 & 5.2080\\
25 & 5.6556 & 6.0590 & 5.6556 & 6.0590 & 6.0383 & 6.0383 & 4.9268 & 5.4180\\
-20 & 4.7472 & 4.6193 & 4.7472 & 4.6193 & 4.8143 & 4.8143 & 4.2552 & 4.0281\\
-21 & 4.9777 & 4.8561 & 4.9777 & 4.8561 & 5.0530 & 5.0530 & 4.4592 & 4.2380\\
-22	& 5.2081 & 5.0929 & 5.2081 & 5.0929 & 5.2919	 & 5.2919 & 4.6633 & 4.4480\\
-23 & 5.4386 & 5.3297 & 5.4386 & 5.3297 & 5.5307	 & 5.5307 & 4.8673 & 4.6579\\
-24 & 5.6691 & 5.5666 & 5.6691 & 5.5666 & 5.7696	 & 5.7696 & 5.0714 & 4.8678\\
-25 & 5.8995 & 5.8035 & 5.8995 & 5.8035 & 6.0085 & 6.0085 & 5.2753 & 5.0777\\
Position(0)/$\mu m$ & 528.3 & -535.3 & 528.3 & -535.3 & -9.7 & -9.7	 & 855.3	 & -810.4\\
Position(5)/$\mu m$ & 501.0 & -516.0 & 619.4 & -654.6 & -29.6 & -139.3 & 842.3 & -819.1\\

\hline\hline
\end{tabular}
\label{tab13}
\end{table*}

\subsection{Effect of focal plane rotation on detection accuracy}

The azimuthal and vertical angles of Ma and the rotation angle of the focal surface change with telescope tacking. Since any of them may cause image size detection error, it is crucial to analyse their effects. Focal plane rotation angle $\theta$ changes with hour angle $t_1$ satisfies \citep{su1997tracking}.

\begin{figure}[htb]
\begin{center}
\includegraphics[width=8cm]{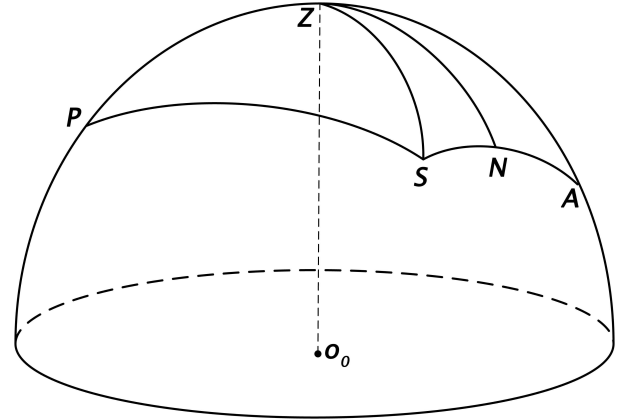}
\caption{Essential points and lines in the celestial sphere. $O_0$ is the spherical centre of the spherical mirror Mb. The optical axis $O_0A$ passes through $O_0$, lies in the meridian plane and is inclined at $25^\circ$ to the horizontal. In a Schmidt system, the centre (vertex) of the reflecting correcting surface of Ma should be at $O_0$, and the surface centre of Mb should be on the optical axis. After reflection by Ma, the ray $SO_0$, which arrives at $O_0$ from the celestial object at the centre of the FOV, passes along the optical axis $O_0A$. The incident plane is the plane that includes the optical axis and the ray $SO_0$. $O_0N$ is the bisector of the angle between the optical axis $O_0A$ and the beam $SO_0$.}
\label{Fig7}
\end{center}
\end{figure}
 
\begin{equation}
	sin\theta=[-cos(A+\phi)+sin(A+\phi)tan(\frac{1}{2}\hat{AS})cosZAS]sin(t_1)
\end{equation}

Where A is $25^\circ$, $\phi$ is the astronomical site latitude, and $\frac{1}{2}\hat{AS}$ is the incident angle of the celestial body, as shown in Fig\ref{Fig7}. Taking the $20^\circ$, $30^\circ$, and $40^\circ$ survey areas as an example, the focal plane rotation angle is shown in the following Table.\ref{tab6}.

\begin{table}[htb]
\caption{Focal plane attitude rotation angle of $20^\circ$, $30^\circ$, and $40^\circ$ survey areas. The rotation angle increases with the hour angle, where the culmination position hour angle is 0. When the hour angle is small ($\leq 3.75$), the rotation angle is small.}
\centering
\begin{tabular}{@{}ccccc@{}}
\hline\hline
$t$ & $\delta=20$ & $\delta=30$ & $\delta=40$ \\
$(^\circ)$ & $(^\circ)$ & $(^\circ)$ & $(^\circ)$ \\
\hline%
$\pm22.5$ & $\mp0.9887$ & $\mp1.2083$ & $\mp3.6119$  \\
$\pm11.25$ & $\mp0.4895$ & $\mp0.5983$ & $\mp1.7889$ \\
$\pm7.5$ & $\mp0.3258$  & $\mp0.3981$ &  $\mp1.1905$ \\
$\pm3.75$ & $\mp0.1627$ & $\mp0.1989$ & $\mp0.5946$ \\
0 & 0 & 0 & 0 \\
\hline\hline
\end{tabular}
\label{tab6}

\medskip
\tabnote{$Notes$: In this table, $t$ is the hour angle and $\delta$ is the survey area.}
\end{table}

The CCD acquisition dimensions of LAMOST are $2k\times 2k$, and the pixel size is $25\mu m$. Based on the focal plane rotation angle, every detector deviation can be calculated. In the $3.75^\circ$ hour angle, the detector deviation is no more than 7 mm, which is far less than detector size. Therefore, focal plane rotation has little effect on the image size.

\subsection{The influence of Ma angle change on spot size detection}

Because of the rotation of the earth, the change of the celestial body angle is $15^{\prime\prime}/s$ as seen by the detector. Mirror Ma needs to control the vertical and azimuthal angles to keep tracking constant; however, this leads to a varying pupil shape and consequently a changing defocus image spot size. Vertical angle is $\angle AZN$; azimuthal angle is $\widehat{ZN}$. Different hour angle and survey area results are shown in Table.\ref{tab7}.

\begin{small}
\begin{table}[htb]
\scriptsize
\caption{Deviation of detected spot size by hour angle in the survey areas of $20^\circ$, $30^\circ$, and $40^\circ$. The results show the deviation increases with the hour angle.(unit: $\mu m$)}
\centering
\label{Tab7}
\setlength{\tabcolsep}{0.6mm}{
\begin{tabular}{@{}ccccccc@{}}
\hline\hline
\multirow{2}{*}{$t$} & \multicolumn{2}{c}{$\delta=20^\circ$} & \multicolumn{2}{c}{$\delta=30^\circ$} & \multicolumn{2}{c}{$\delta=40^\circ$} \\
&  Horizontal &  Vertical    
&  Horizontal &  Vertical 
&  Horizontal &  Vertical   \\
\hline
$\pm22.5^\circ$   & -193.9 & 44.0 & -218.8 & 50.9 & -241.6 & 56.7 \\
$\pm11.25^\circ$  & -44.9  & 10.9 & -56.5  & 13.1 & -62.6  & 14.6 \\
$\pm7.5^\circ$    & -22.3  & 4.9  & -25.3  & 5.8  & -27.9  & 6.5  \\
$\pm3.75^\circ$   & -5.6   & 1.2  & -6.3   & 1.5  & -7.0   & 1.6  \\
0 & 0 & 0 & 0 & 0 & 0 & 0 \\
\hline\hline
\end{tabular}}
\label{tab7}
\end{table}
\end{small}

The result reveals that the experimental observation time needs to be controlled to within 20 minutes and that a celestial body observed near the culmination should be selected.

\section{Experiment and discussion}

Before detection of the focal plane attitude of LAMOST, it is necessary to calibrate other parts of the optical path. The most important is coaxial detection. The tilt of Ma and Mb can be calibrated using active optics technology before each observation.

Coaxial detection requires a theodolite and specified marks in the optical path, which include the Shack-Hartmann centre(SHC) of the Ma mirror, two cross marks at the top (TS) and bottom (BS) of the slope between Ma and Mb, the centre of the Mb (CMB) mirror, and cross marks on the top (TF) and bottom (TB) of the focal plane.  The centre of Mb is marked by pasting the reflector.

The theodolite is set on the line between the SHC of the Ma mirror and the cross mark of the bottom slope. Then, the azimuth angle and vertical angle of the marks is detected. 

\begin{table}[htb]
\caption{Coaxial detection of LAMOST. The standard azimuth angle value of SHC is $180^\circ$. The standard azimuth angle of the other marks is $0^\circ$. The standard azimuth angle between SHC and CMB is $-25^\circ$.}
\centering
\begin{tabular}{@{}ccccc@{}}
\hline\hline
Detection position & Azimuth angle & vertical angle \\
\hline%
SHC & $180^\circ0^\prime13.3^{\prime\prime}$ & $-25^\circ0^\prime0.2^{\prime\prime}$ \\
TS & $-0^\circ0^\prime2.3^{\prime\prime}$ & $-12^\circ25^\prime18.9^{\prime\prime}$ \\
BS & $-0^\circ0^\prime1,7^{\prime\prime}$ & $-4^\circ3^\prime25.1^{\prime\prime}$ \\
CMB & $0^\circ0^\prime11^{\prime\prime}$ & $25^\circ0^\prime3^{\prime\prime}$ \\
TF & $-0^\circ0^\prime14.8^{\prime\prime}$ & $-25^\circ0^\prime12.1^{\prime\prime} $ \\
BF & $-0^\circ0^\prime57.2^{\prime\prime}$ & $-25^\circ0^\prime35.4^{\prime\prime} $  \\
\hline\hline
\end{tabular}
\label{tab8}
\end{table}

The theodolite is set on the line between the SHC of the Ma and the cross mark of the bottom slope. The result is shown in the Table.\ref{tab8}.

The results of the coaxial detection show that the Ma, Mb, and focal panel coaxial detection accuracy are within 13 arcseconds, and have less effect on focus surface attitude detection.

At the start of observation or before changing the survey area: When the target central star is near the culmination, we adjust the telescope focal plane position from positive direction defocus (toward Ma) to negative direction defocus  (toward  Mb). Imaging in the 20, 21, 22, 23, 24, 25, -20, -21, -22, -23, -24, -25 mm positions respectively, eight acquisition cameras are used to capture the image. The duration of the experiment was limited to within 20 minutes to reduce the effect of star position on the size of the image.

\begin{figure}[htb]
\begin{center}
\includegraphics[width=8cm]{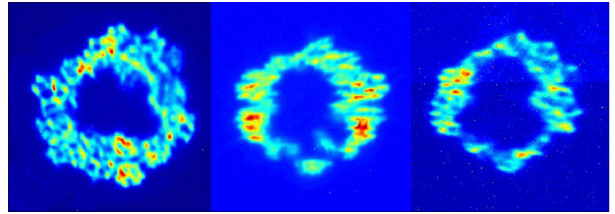}
\caption{Images from three cameras are presented. Left: No.0 camera. Center: No.4 camera. Right:  No.6 camera. The central dark spots are caused by shading of the focal plane and the central segmented mirror of Ma, which is masked.}
\label{Fig8}
\end{center}
\end{figure}

Although LAMOST is equipped with eight acquisition cameras that produce eight sets of data, it was a challenge to find multiple nonoverlapping bright stars for detection testing and LAMOST tracking requirement only guarantee four cameras have a bright star. In the March 2019 test, we were able to obtain three suitable targets. The results of image size and image detection are shown in the Table.\ref{tab9} and Fig.\ref{Fig8}.

\begin{table}[htb]
\caption{The detection results of three CCD cameras and defocus distance has show good linearlylinearity. Using the look-up table, we give the three-position focal point deviation.}
\centering
\begin{tabular}{@{}ccccc@{}}
\hline\hline
Defocus distance & 0 CCD & 4 CCD & 6 CCD \\
\hline%
-25 & 235.098 & 195.978 & 202.500 \\
-24 & 224.963 & 187.634 & 193.336 \\
-23 & 215.082 & 182.261 & 184.632 \\
-22 & 204.364 & 173.729 & 176.664 \\
-21 & 192.835 & 166.906 & 166.464 \\
-20 & 185.727 & 158.908 & 158.835 \\
20 & 179.343 & 164.032 & 168.807 \\
21 & 188.298 & 172.645 & 176.893 \\
22 & 198.329 & 180.548 & 186.920 \\
23 & 209.109 & 189.304 & 194.478 \\
24 & 219.508 & 196.662 & 203.660 \\
25 & 230.452 & 205.187 & 211.806 \\
focal position & 440.4 & 855.0 & -716.8 \\
\hline\hline
\end{tabular}
\label{tab9}
\end{table}

The No.0 detector corresponds to the -1.5/1.5 FOV simulation, while the fourth detector corresponds to the 0/2.5 FOV simulation and the sixth detector corresponds to the 0/-2.5 FOV simulation.

A right-handed coordinate system is constructed, in which the direction from Mirror Mb to Ma is set as the  positive direction of the Z-axis, horizon direction is set as X-axis, vertical direction is set as Y-axis. Comparing the experiment data with the look-up table, the test results show that the No.0, No.4 and No.6 detectors have deviations of $-179\mu m \sim -87.9\mu m$, $12.7\mu m$, and $102.3\mu m$, respectively. According to the geometric relationship, there is a reverse 56.2 $\sim$ 97.9 arc-second tilt around the horizontal axis and a positive 10.2 arc-second tilt around the vertical axis.

\section{Conclusion}

In this study, we used eight acquisition cameras to detect the focal plane attitude of LAMOST. While applying the focal plane defocus imaging method to detect focal plane attitude, we considered the influence of Ma and focal plane rotation and the collimation of the optical path during the tracking process. By limiting the experimental observation time and detecting near the culmination, we effectively controlled the influence of Ma and focal plane rotation.

The estimated focus position obtained by our method deviated from the focal surface. To correct the deviation, we used a look-up table method. The focus position detection accuracy achieved by the simulation experiment reached 30 arcsec.

Before detection of the focal plane attitude of LAMOST in our experiment, we calibrated the collimation of the optical path using a theodolite. In addition, we calibrated the tilt of the Ma and Mb mirrors by active optics to reduce the influence of tilt of Ma and Mb on the attitude detection of the focal plane.

According to the experimental results, According to the geometric relationship, there is a reverse 56.2 $\sim$ 97.9 arc-second tilt around the horizontal axis and a positive 10.2 arc-second tilt around the vertical axis.

Additional experimental observations are required to verify the reliability of our method. Advancing our method aims to decrease the effect of attitude detection on normal observations. At the same time, the method can be applied to focal plane attitude detection in other large-field telescopes.

\begin{acknowledgements}
We would like to express our gratitude to the LAMOST work group and for the help of VISTA's Gavin Dalton. 

This work is supported by the National Nature Science Foundation of China Grant U1931207. The Guo Shou Jing Telescope (or LAMOST) is a National Major Scientific Project built by the Chinese Academy of Sciences. Funding for the project has been provided by the National Development and Reform Commission. LAMOST is operated and managed by National Astronomical Observatories, Chinese Academy of Sciences. 

\end{acknowledgements}

%
%
%

\bibliographystyle{raa-mnras}
\bibliography{raa_paper_notes}

\end{document}